\documentclass{article}

\usepackage{arxiv}

\usepackage[dvipsnames]{xcolor}
\usepackage[utf8]{inputenc} 
\usepackage[T1]{fontenc}    
\usepackage[colorlinks=true,linkcolor=Magenta,citecolor=black,urlcolor=Magenta]{hyperref}       
\usepackage{url}            
\usepackage{booktabs}       
\usepackage{amsfonts}       
\usepackage{nicefrac}       
\usepackage{microtype}      
\usepackage{xspace}
\usepackage{float}
\usepackage[nolist]{acronym}
\usepackage{pdflscape}
\usepackage{enumitem}

\begin{acronym}
\acro{CTF}{Capture the Flag}
\acro{AFNOM}{AFiniteNumberOfMonkeys}
\acro{DL}{Distance Learning}
\acro{WTCTF}{WhatTheCTF?!}
\acro{PBL}{Problem-Based Learning}
\acro{PjBL}{Project-Based Learning}
\acro{HTM}{HackTheMidlands}
\end{acronym}

\usepackage{tikz}

\usepackage[square,numbers]{natbib}

\newcommand{\etal}{\textit{et.\ al.}\xspace}

\title{Learn-Apply-Reinforce/Share Learning: Hackathons and \acsp{CTF} as General Pedagogic Tools in Higher Education, and Their Applicability to Distance Learning}

\author{
  Tom Goodman$^{\dagger}$ and Andreea-Ina Radu$^{\ddagger}$ \\
  Department of Computer Science\\
  University of Birmingham\\
  United Kingdom, B15 2TT \\
 $^{\dagger}$\texttt{t.a.goodman@cs.bham.ac.uk}, $^{\ddagger}$\texttt{a.i.radu@cs.bham.ac.uk}
}

\begin{document}
\maketitle

\begin{abstract}
\textbf{This paper lays out two teaching/learning methods that are becoming
  increasingly prevalent in computer science - hackathons, and \ac{CTF}
  competitions - and the pedagogic theory that underpins them. A case study of
  each is analysed, and the underpinning similarities extracted. The frameworks
  are then generalised to Learn-Apply-Reinforce/Share Learning - a social
  constructivistic method that can be used subject-independently. The
  applicability of this new method to distance learning is then investigated -
  with a mind to potential necessity to work from home - both due to increasing demand in
  the Higher Education sector, but also the devastating impact of crises such as 
  the ongoing COVID-19 pandemic. Finally, a few potential extensions and future
  applications are discussed - including the possibilities of pivoting the method
  to be more research-driven, or indeed, to drive research.}

\end{abstract}


\section{Introduction} \label{sec:introduction}

Hackathons and \acp{CTF} have long been a staple of Computer Science, with the
first software hackathon being held by Sun Microsystems at their JavaOne
conference in June 1999 \citep{aviram_1999}, and the notion of security
\acp{CTF} (or wargames) being around since time immemorial. Over time, they
have become increasingly popular, with significant uptake not only by
students\footnote{\url{https://mlh.io/about}}\textsuperscript{,}\footnote{\url{https://ctftime.org/ctfs}},
but also in industry. Not only do these constructs offer invaluable frameworks
for \ac{PBL}, they also pose a potential solution to inspiring
and attracting a diverse range of applicants to Higher Education - not least in
subjects like computer science where this is especially pertinent
\citep{main2017underrepresentation}. 

With the rise of distance learning in Higher Education institutions - both due
to overall demand, and situations such as the ongoing COVID-19 crisis - it is
becoming increasingly necessary to consider a wide variety of innovative and
engaging methods for delivering teaching. This paper, therefore, delves deeper
into the essence of both hackathons and \acp{CTF}, and aims to generalise them
such that they could be applied in a non-discipline-specific way. Furthermore, it
is believed that these approaches could be altered further to incorporate a
more research-driven angle - making them of particular value to
research-intensive institutions. 

\subsection{Hackathons}
\begin{figure}
    \centering
    \includegraphics[width=0.75\linewidth]{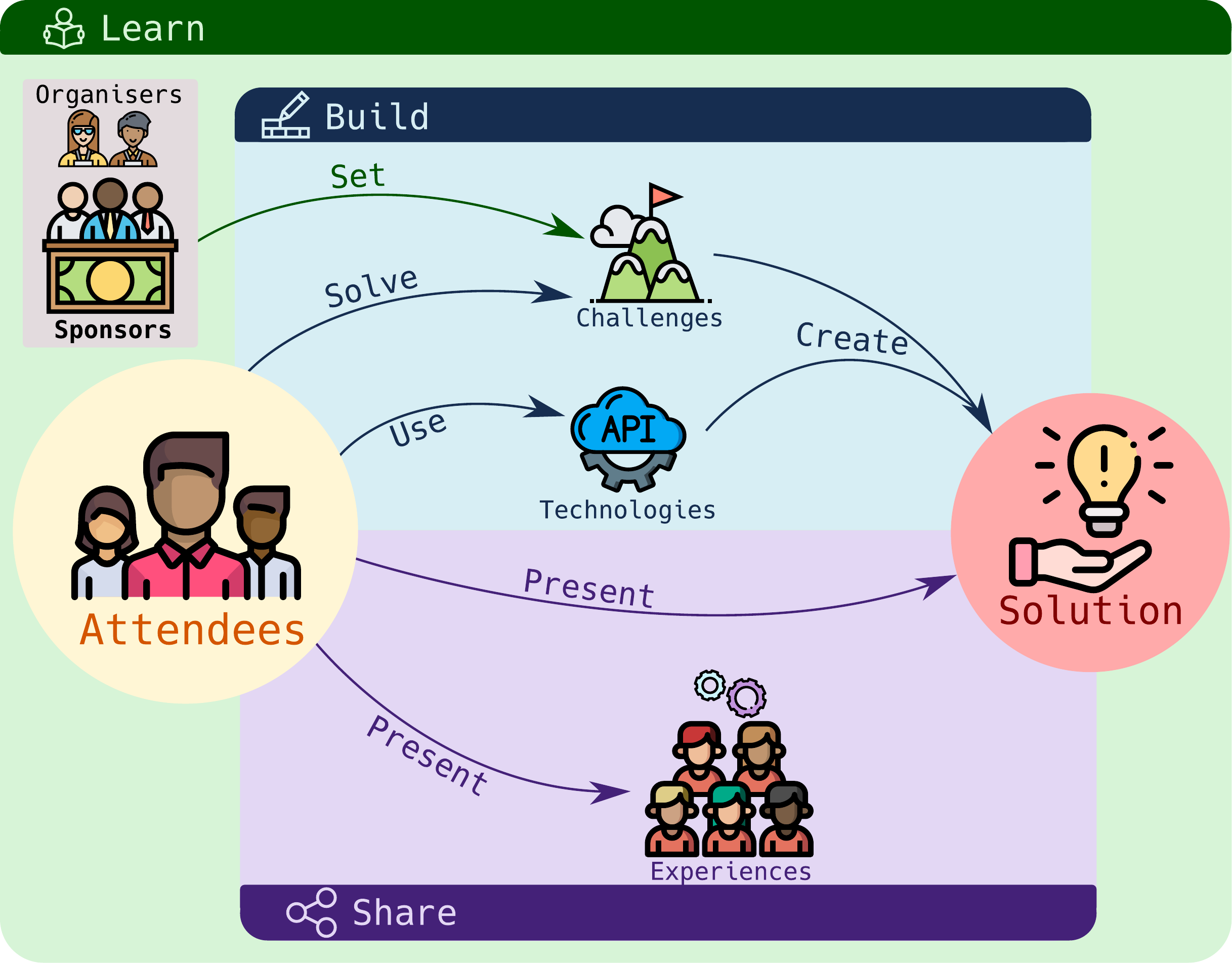}
    \caption{Diagram showing the overall framework of a general hackathon.
    Interactions between entities are depicted as arrows.}
    \label{fig:hackathon}
\end{figure}

Hackathons are `creative marathons' where people (in the context of this paper,
students) come together in teams to collaborate - learning new ideas, creating
something innovative or novel, and sharing these inventions to the other
participants. By pulling together three tenets (learn, build, and share), they
foster a truly unique learning environment where participants build up not only
their hard skills, but also their soft skills (i.e.\ through presentation,
communication, etc.). Traditionally participants work (usually in teams of 4)
towards broad challenges set by sponsors (often companies) such as `the project
that could do the most good for society', or `the best use of \textit{some
specific technology}. As discussed in Section~\ref{sec:generalisation}, the
framework provided by hackathons (Figure~\ref{fig:hackathon}) can be generalised
(along with that of \acp{CTF}) to teaching in Higher Education on the whole.

Hackathons are clear examples of \ac{PBL}
\citep{savery1995problem} -- with sponsors' challenges broadly acting as the
problems being solved, and reinforced by the various technical problems faced
in the conception and building of solutions. This teaches not only the content
(in this case, about the `challenges' set), but also about the thinking
strategies and technologies that can be used when solving them (and similar
problems encountered in the future) \citep{hmelo2004problem}. Further, the
collaborative, community-focused environment often fostered at hackathons
encourages significant peer mentoring - even between `competing' teams
\citep{lara2016hackathons}. By virtue of their underpinning nature, hackathons
additionally exhibit features of \ac{PjBL}
\citep{blumenfield1991problem, kokotsaki2016problem}. By bringing
together such a diverse range of learning vectors, hackathons are able to
stimulate curiosity in learners, and engage diverse ranges of learners. 

\subsection{\acp{CTF}} 
Historically, Capture the Flag competitions have existed within the context of
cyber security, whereby teams\footnote{Generally, \acp{CTF} are played in
teams. However, individual \acp{CTF}s exist as well, for example, 
as qualifiers for an in-person finale (e.g.
\href{https://cambridge2cambridge.csail.mit.edu/2017_event}{C2C 2017}).} play
against each other to obtain as many \textit{flags} as possible, by solving
\textit{challenges}, set by the organisers. As an analogy to traditional
teaching, the challenges can be viewed as exercises or assignments set by
teachers or lecturers. There are two flavours of \ac{CTF} - attack/defence, or
jeopardy style - with the latter being the most popular.  In an attack/defence
CTF, teams have to protect a (purposefully vulnerable) host, while at the same
time attacking other teams' hosts, which have the same vulnerabilities. In a
jeopardy CTF, participants solve a number of stand-alone challenges, across a
range of categories (e.g.\ web, cryptography, binary exploitation, reverse
engineering, forensics, networking, etc).

\acp{CTF} illustrate best the combination of \ac{PBL} and online learning. Most
\acp{CTF} are held online\footnote{Some \acp{CTF} hold online qualifiers, and
in-person finales. The online stage is open to \textit{everyone}, while the top
\textit{n} teams are invited for the finale, where team sizes are limited, and
some financial support is provided for transport and accommodation.}, and teams
can be made up of individuals from across the world, thereby shattering any
geographical restrictions in-person events have. In terms of learning outcomes,
\acp{CTF} provide a good opportunity for participants to sharpen existing
skills by focusing on categories they are familiar with, or acquire new
knowledge by engaging with new categories, as challenges vary in difficulty
levels within categories. Solving \ac{CTF} challenges means exploiting
vulnerabilities in software/hardware and, therefore, this type of competition
has a particularly close relationship with the state-of-the-art in the cyber
security field. Furthermore, \acp{CTF} informally address some of the key
factors identified by the Government in the National Cyber Security Strategy
2016-2021: the lack of young people entering the profession, and insufficient
exposure to cyber and information security concepts in computing
courses~\citep{hammond2016national}.

\begin{figure}
  \centering
  \includegraphics[width=0.785\linewidth]{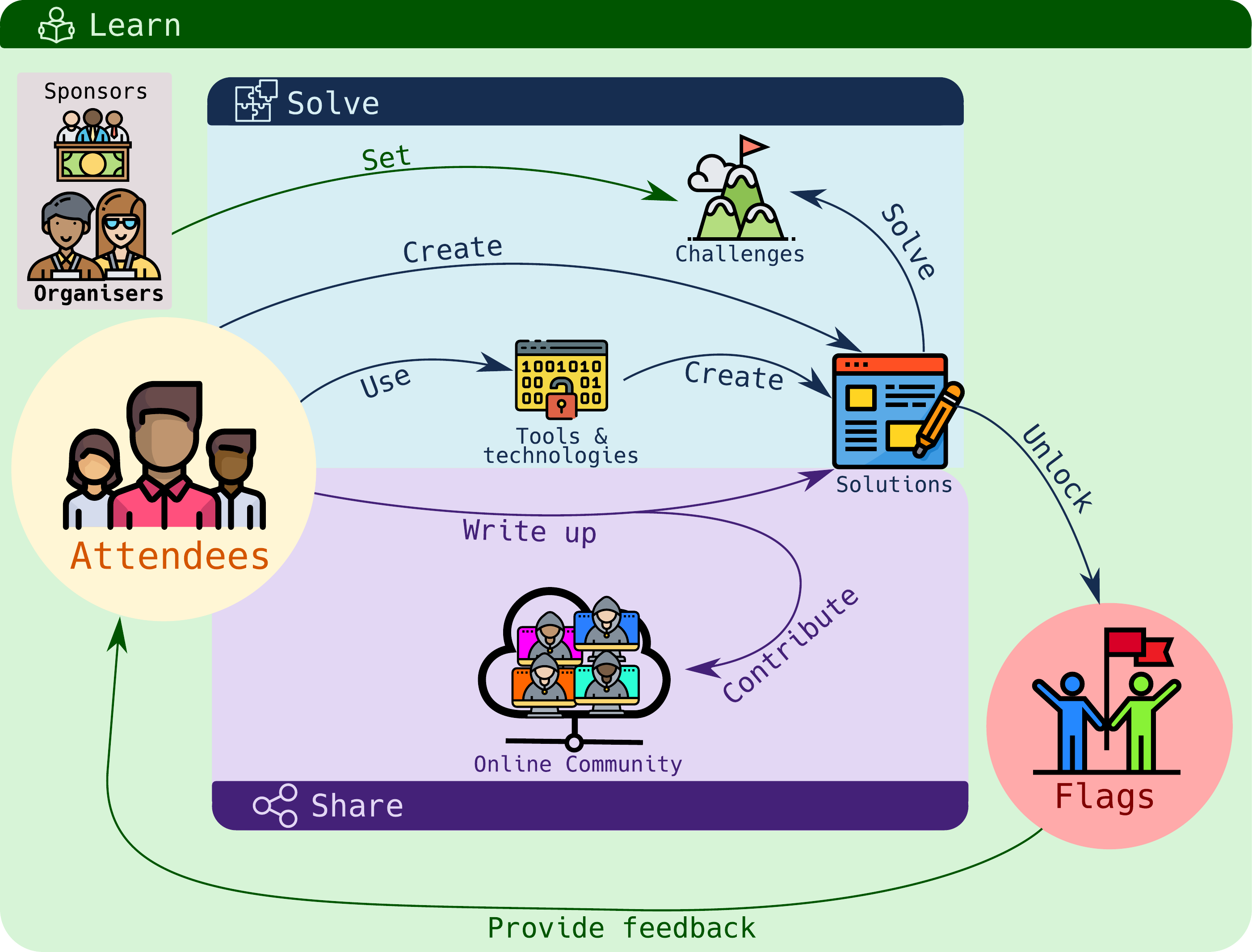}
  \caption{Diagram showing the overall framework of a general \ac{CTF}.
    Interactions between entities are depicted as arrows.}
  \label{fig:ctf}
\end{figure}

Figure~\ref{fig:ctf} presents an overall framework of a \ac{CTF}. Organisers
(and sometimes sponsors) create a set of challenges (prior to the CTF itself), which teams then work
together to solve. Various tools and technologies can aid participants in
solving the challenges (e.g.\ the Kali
Linux\footnote{\url{https://www.kali.org/}} operating system comes preloaded
with an extensive suite of software useful for \acp{CTF}). Some challenges
require programming skills, while others can be solved by conventional methods
(e.g.\ cryptography challenges could require pen-and-paper solutions). 
Solving a challenge yields a flag, which is then submitted to a scoreboard
(omitted from diagram for simplicity purposes), in return for points. This allows
participants to have immediate feedback on the correctness of their solution.
After the competition ends, teams often write up their solutions for challenges
and share them with the wider online community. Learning continues through
participants being able to look up challenges which they have not have been 
able to solve during the \ac{CTF}, and learn from the write-ups of their
peers.

While \acp{CTF} are generally short events, (they last from a few hours to
\textasciitilde3 days), they have also been adapted and used as tools for
teaching and learning in Higher Education, over greater periods of time.
Mirkovic and Peterson~\citep{mirkovic2014class} describe how to adapt \acp{CTF}
to `Class-\acp{CTF}', classroom-based exercises, such that learning is done
over a longer period of time (over the length of a course) and all students
reach their learning objectives.  Flushman~\etal~\citep{flushman2015not} have
developed a full introductory computer science course for first-year
undergraduates, based on \acp{CTF} and linear alternate reality games.
Chothia~\etal~\citep{chothia2019choose} explain how they used CTF-style
exercises for an 11-weeks cyber security course, adding a story throughout the
assignments, which increased engagement and attainment levels in students.
\acp{CTF} have also been used at Further Education level (high-school), in
order to try and address the shortage of students who are interested in
studying cyber security during their university
studies~\citep{feng2016divergent}. Moreover, a number of online learning
platforms~\citep{htb,immersive,tryhackme} have appeared over the last years,
generally aimed at being `training grounds' for \acp{CTF}. These platforms
could also be used as resources for \ac{DL}.


\section{\acl{HTM}} \label{sec:htm}

\begin{figure}
    \centering
    \includegraphics[scale=0.19]{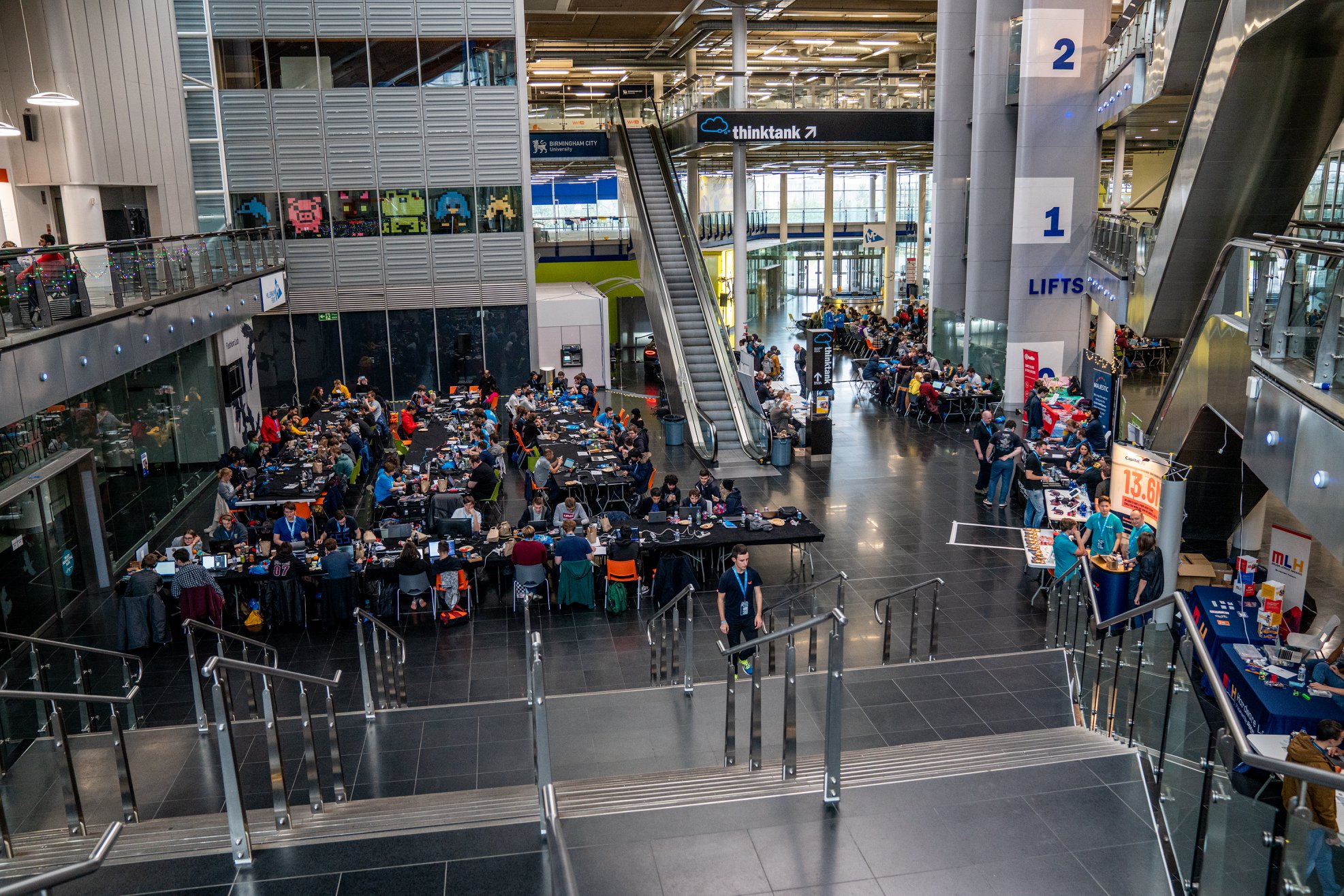}
    \caption{A shot from \acl{HTM} 4.0 showing some of the scale - both
    of sponsors and participants at the event.}
    \label{fig:htm_pic}
\end{figure}

\begin{figure}
\begin{minipage}{0.47\textwidth}
    \includegraphics[width=\linewidth]{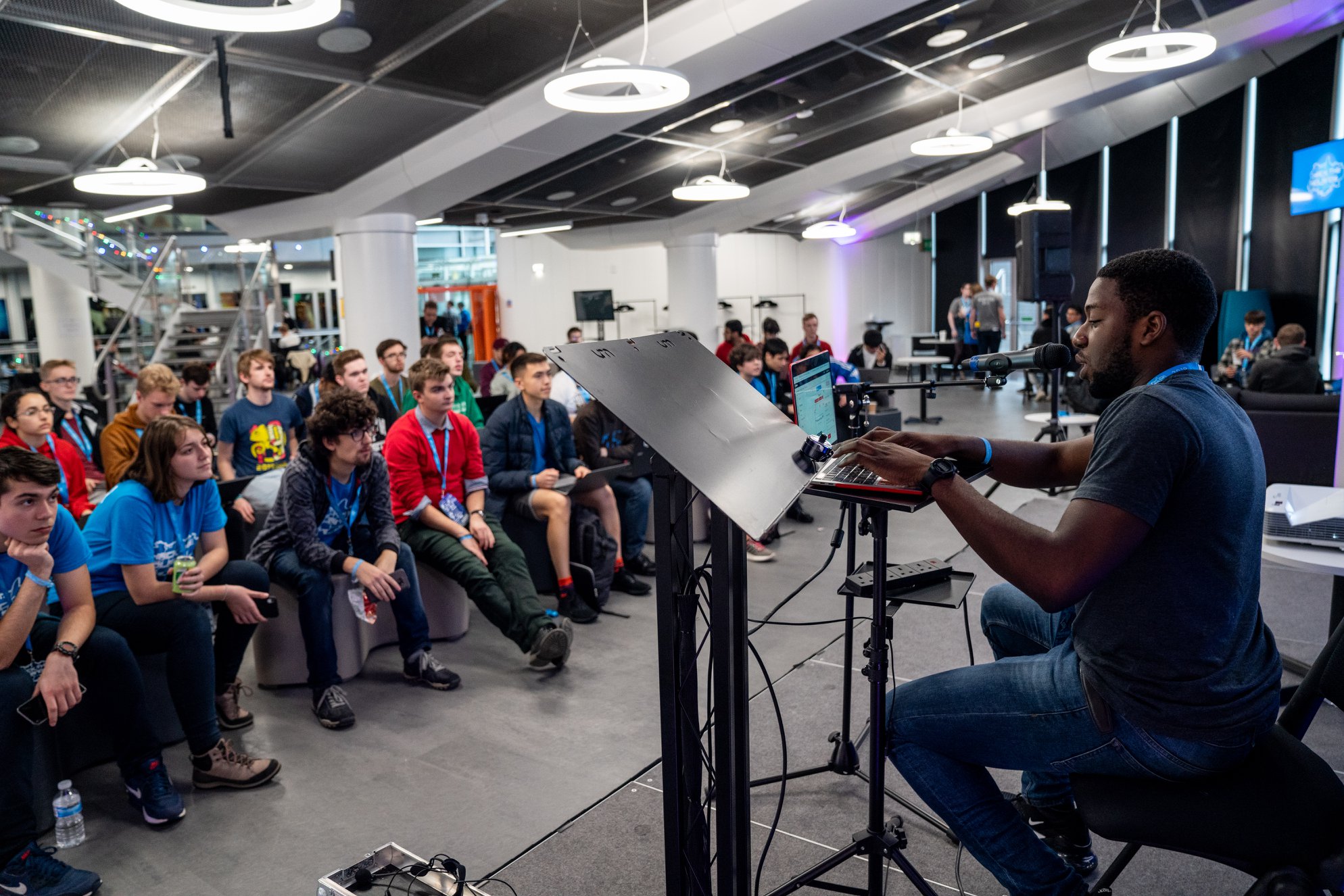}
    \end{minipage}
    \hspace{\fill} 
    \begin{minipage}{0.47\textwidth}
    \includegraphics[width=\linewidth]{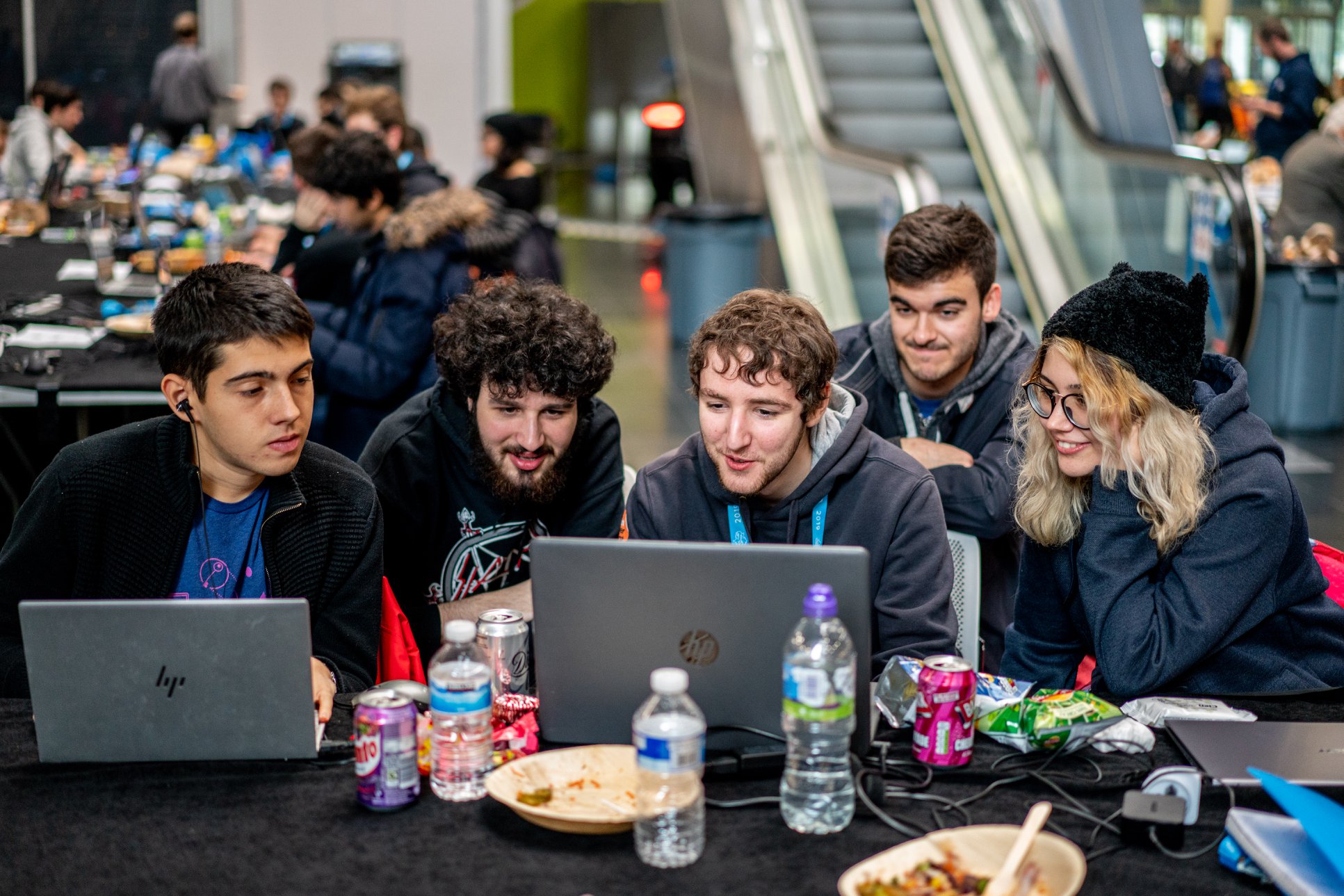}
    \end{minipage}

    \vspace*{0.5cm} 

    \centering
    \begin{minipage}{0.6\textwidth}
    \includegraphics[width=\linewidth]{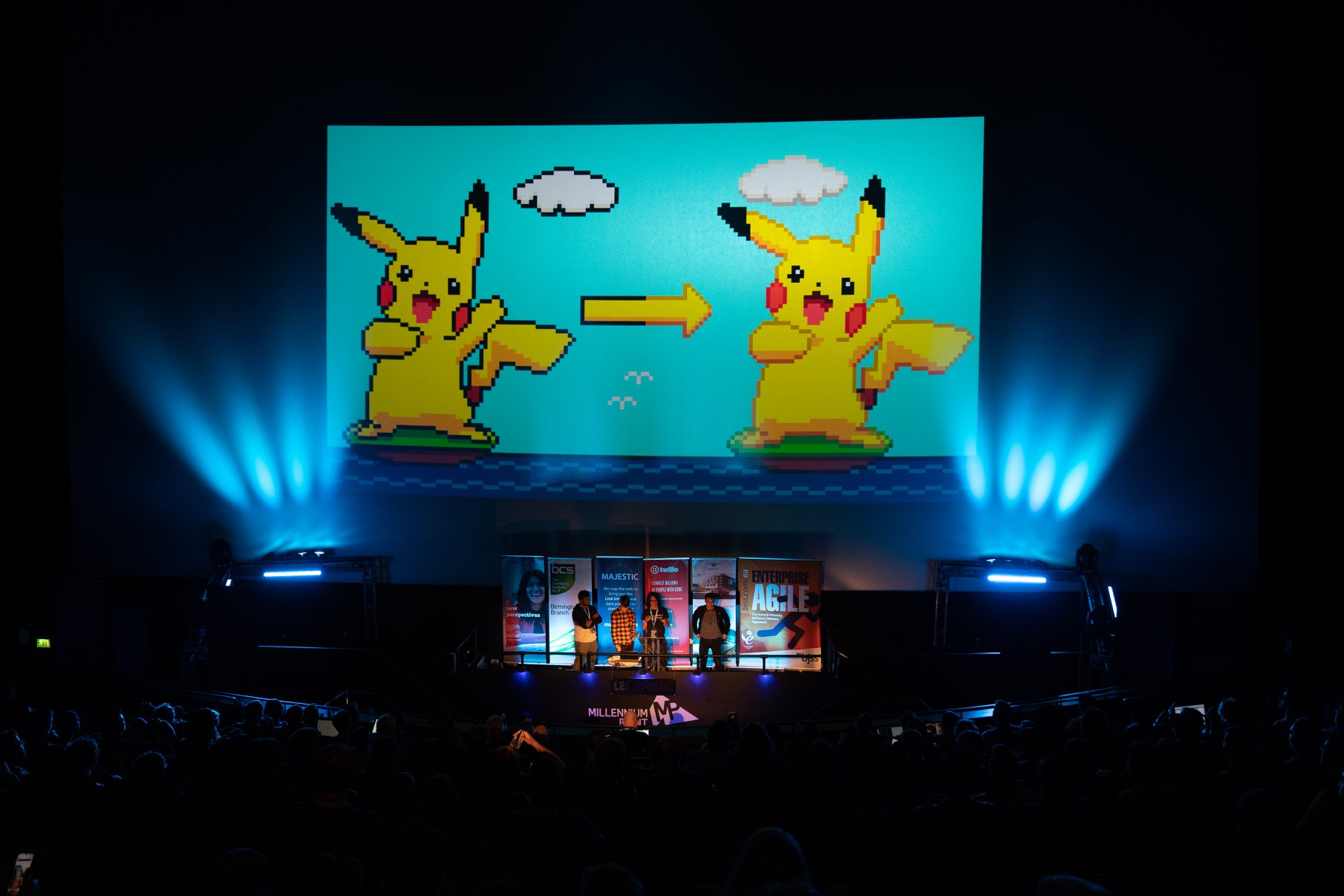}
    \end{minipage}
  \caption{\textbf{Top Left}: Learn - a small-scale workshop where attendees
  are learning about the Twilio API; \textbf{Top Right}: Build - a team of
  attendees working hard on their project; \textbf{Bottom}: Share - a team of
  four showing off their project (an automatic selective outliner for pixel art)
  on an IMAX screen to over 300 people.} \label{fig:learnbuildshare}
\end{figure}



\ac{HTM}\footnote{\url{https://hackthemidlands.com/}} is a yearly-hosted
hackathon in Birmingham, UK\@. Founded in 2016, the distinguishing factor from
most other hackathons (particularly MLH-partnered ones) is that it is open to
everyone. This includes non-students of all ages, and under-18s. This diversity
encourages vastly wider interaction between traditionally distanced groups, and
facilitates the sharing of a wide range of experiences, knowledge, and
opinions, and promotes unlikely collaborations. This diversity of makeup helps
to foster creativity and innovation in much the same way Bercovitz and Feldman
observed of researchers \citep{bercovitz2011mechanisms}. 

Operationally, \ac{HTM} functions in much the same way as other
hackathons. Attendees congregate in a single physical location (allowing a
strong community feel to be fostered), and work together over the span of
\textasciitilde 24 hours to `hack together' solutions to often broad and
real-world challenges or problems set by the sponsors (Figure~\ref{fig:htm_pic}). Generally the sponsors
are companies (large or small), but they have also been more charitable or
academic institutions such as the University of Birmingham or the
BCS\footnote{\url{https://bcs.org/}}. In addition, \ac{HTM} partners
with a number of organisations each year - especially to improve outreach, such
as work with Arkwright\footnote{\url{https://arkwright.org.uk/}} and the Social
Mobility Foundation\footnote{\url{https://www.socialmobility.org.uk/}}.

On the first of the two days (preceded by the opening ceremony, in which
sponsors pose their challenges) the organisers put on what amounts to a
mini-conference - two tracks (one entry-level, one more advanced) of talks and
workshops to equip attendees with relevant skills. In addition to this, a
mentor scheme - with mentors consisting both of previous attendees and industry
experts - helps attendees to access help with little preventative barrier.
Whilst the mentor scheme and workshops support the `learn' and `build' aspects
of the hackathon, the `share' part is predominantly propped up by the volunteer
and organiser teams (Figure~\ref{fig:learnbuildshare}). They actively encourage all attendees to not only present
what they have made, but also what they have learned, and the experience they
have had. This in turn means that all learners are reassured that they have
learned or built something valuable, which is worth sharing. Given that
presenting in front of a 300+-strong crowd is daunting at the best of times,
these reassurances lead to a significant proportion of participants giving
presentations. 

\begin{figure}
    \centering
    \includegraphics[scale=0.375]{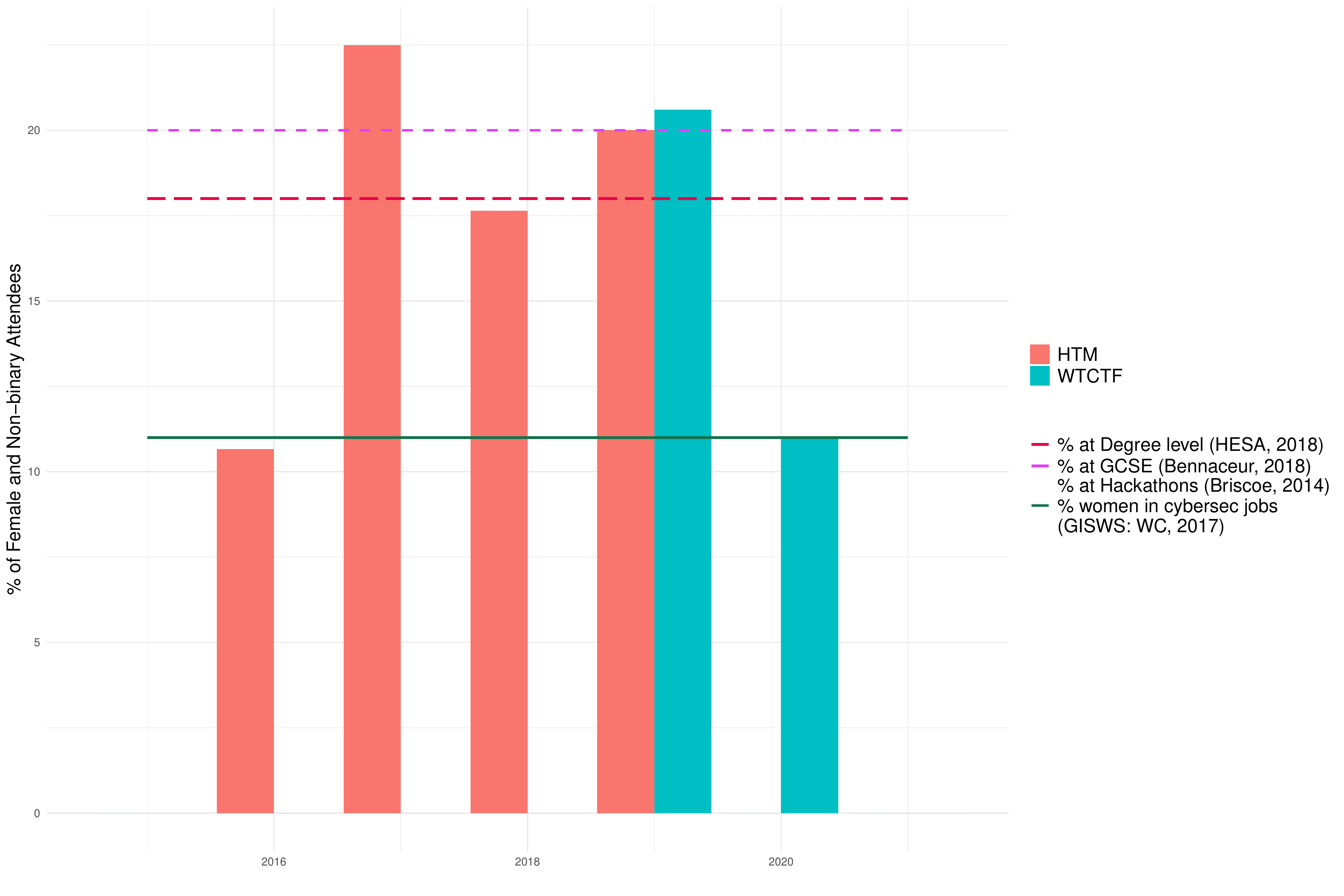}
    \caption{Graph showing female and non-binary attendance of \acl{HTM} 
    and \acl{WTCTF} compared to general levels set out by \citep{briscoe2014digital}, \citep{hesa}, and \citep{bennaceur2018diversity}.}
    \label{fig:gender}
\end{figure}

Further, \ac{HTM} has made a concerted effort to be an informal (as
described by Nandi and Mandernach \citep{nandi2016hackathons}), and
approachable learning environment. This not only encourages a diverse range of
participants (especially from disadvantaged or underrepresented groups) to take
part, but also engages with those in Higher Education who are typically
disinterested by traditional teaching environments such as lectures. In
addition, by leveraging the concept of delight \citep{kim2015relationships}
(simply put, going above and beyond for the learner), it is possible to augment
the learning experience by making the students more conducive to learning. This
has the additional effect of creating a positive and supportive culture at the
event, which leads to peer mentoring, and a willingness to ask for help more
readily. This not only improved the overall learning that participants
experience, but also helps to improve diversity at the event - a particularly
clear issue in computer science. In comparison to the figures set out by
Briscoe \citep{briscoe2014digital}, HESA \citep{hesa}, and Bennaceur \etal
\citep{bennaceur2018diversity} \ac{HTM} reports a higher proportion of
female and non-binary attendees in all but their first year - likely down to
the additional steps they took to reach out to and include underrepresented
groups (Figure~\ref{fig:gender}).

Participant feedback is at the core of \ac{HTM}, as it is imperative to
act on it to upkeep delight and satisfaction (and thus, maintain a high quality
of teaching). To this end, \ac{HTM} surveyed attendees post-event to
ascertain what was positive and negative about the experience, and moreover,
what could be improved on the whole. Some of these questions were shared with
`\acl{WTCTF}' in order to facilitate comparisons between the two
(Section~\ref{sec:comparison}).

\section{\acl{WTCTF}} \label{sec:wtctf}
\ac{WTCTF}\footnote{\url{https://wtctf.afnom.net}} is a yearly-hosted \ac{CTF} 
at the University of Birmingham. Its audience is predominantly computer science students,
however the event is open to all students of the university. The first iteration
of \ac{WTCTF} was in 2019, when a group of members of the Ethical Hacking Club
\ac{AFNOM}\footnote{\url{https://afnom.net}}, having experienced a number of
\acp{CTF}~\cite{radu2015organising}, decided it was an opportune time to
organise such an event for the University. They set out with the goal of
creating a beginner-friendly \ac{CTF}, such that all students can successfully
participate, learn new skills and build up their confidence in an area they
might not know much about. 

One of the aims was to entice people that did not generally have an
interest in cyber security, with the hope that they would start considering
such a path for their studies or career afterwards. The set of challenges
created covered a wide range of topics: web, cryptography, forensics, binary
reverse engineering and exploitation, forensics, steganography, and password
cracking. Participants could approach the challenges in a `horizontal' way,
trying out a small number of challenges from all categories, in order to form a
broad idea of what topics cyber security covers, or could use a `vertical'
approach, diving deep within one or two categories. In each category, the
difficulty of the challenges was signalled by the amount of \textit{points}
assigned to them. Some challenges employed the \textit{scaffolding} method of
learning, whereby solving one challenge unlocked a new, very similar one, with
a slightly increased difficulty.

\ac{WTCTF} ran over seven hours, with a break for lunch, when all challenges
were unreachable to participants. This encouraged people to network, exchange
ideas and, simply take a break from the grip of the competition. In terms of
attendees, \ac{WTCTF} had a lower number of participants (\textasciitilde35) than \ac{HTM},
both due to the specialised nature of the event, as well as being open only to
University of Birmingham students. Figure~\ref{fig:gender} shows the percentage of
female attendees, as compared to the percentage of women following a cyber
security career~\citep{reed20172017} and percentage of women currently studying
Computer Science at degree level~\citep{hesa}.

%
%
%

During \ac{WTCTF}, the main learning vector is \textit{collaborative learning},
done within teams, and significant amounts of self-directed learning from the
participants is also required. The learning and solving components, as described 
in the general CTF framework (Figure~\ref{fig:ctf}) are tightly coupled together,
and happen simultaneously. Studies have 
shown that collaborative learning
improves learning outcomes for students (participants), with respect to academic
achievement, quality of inter-personal interactions, self-esteem and perceived
social
support~\citep{prince2004does,johnson1998cooperative,springer1999effects}.
Moreover, collaborative learning appears to improve retention of traditionally
under-represented
groups~\citep{fredericksen1998minority,berry1991collaborative}. One of the soft
skills most improved by \ac{WTCTF} are communication skills, as the time limited
nature of the event means information must flow quickly and accurately between
teammates.  Furthermore, the organisers provide direct support during the
competition, as participants can just walk up to them and ask questions about
challenges.  Organisers can choose to provide hints or point participants to
specific resources. Due to the smaller scale of the event, organisers can also
closely monitor teams, identify any that are struggling, and provide additional
help.  One of the aims of \ac{WTCTF} is to make sure newcomers gain new skills in cyber security, and the ability to provide close support is core to achieving this. 

\begin{figure}
\centering
\begin{minipage}{0.5\textwidth}
    \includegraphics[width=\linewidth]{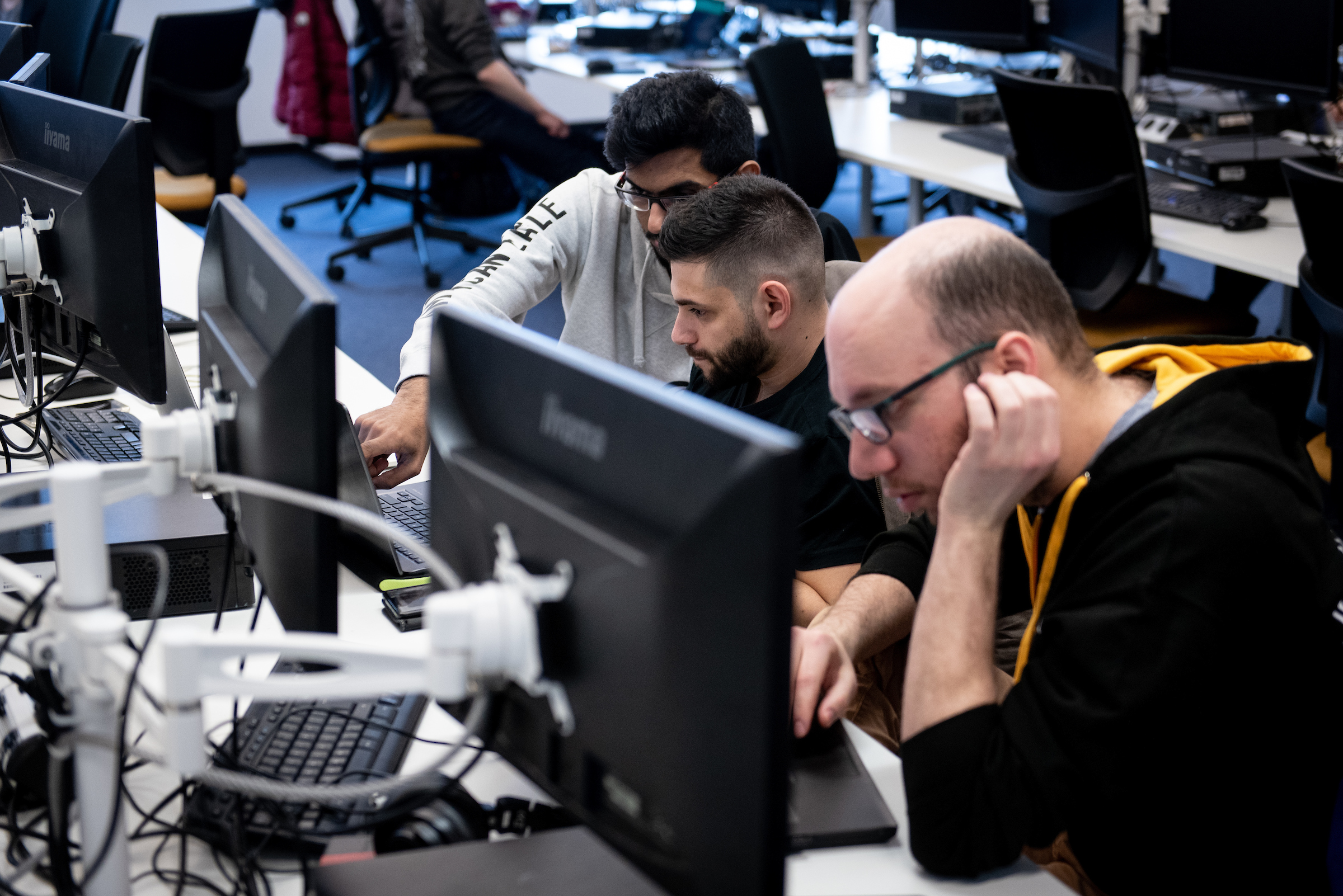}
    \end{minipage}
    \hspace{\fill} 
    \begin{minipage}{0.47\textwidth}
    \centering
    \includegraphics[width=\linewidth]{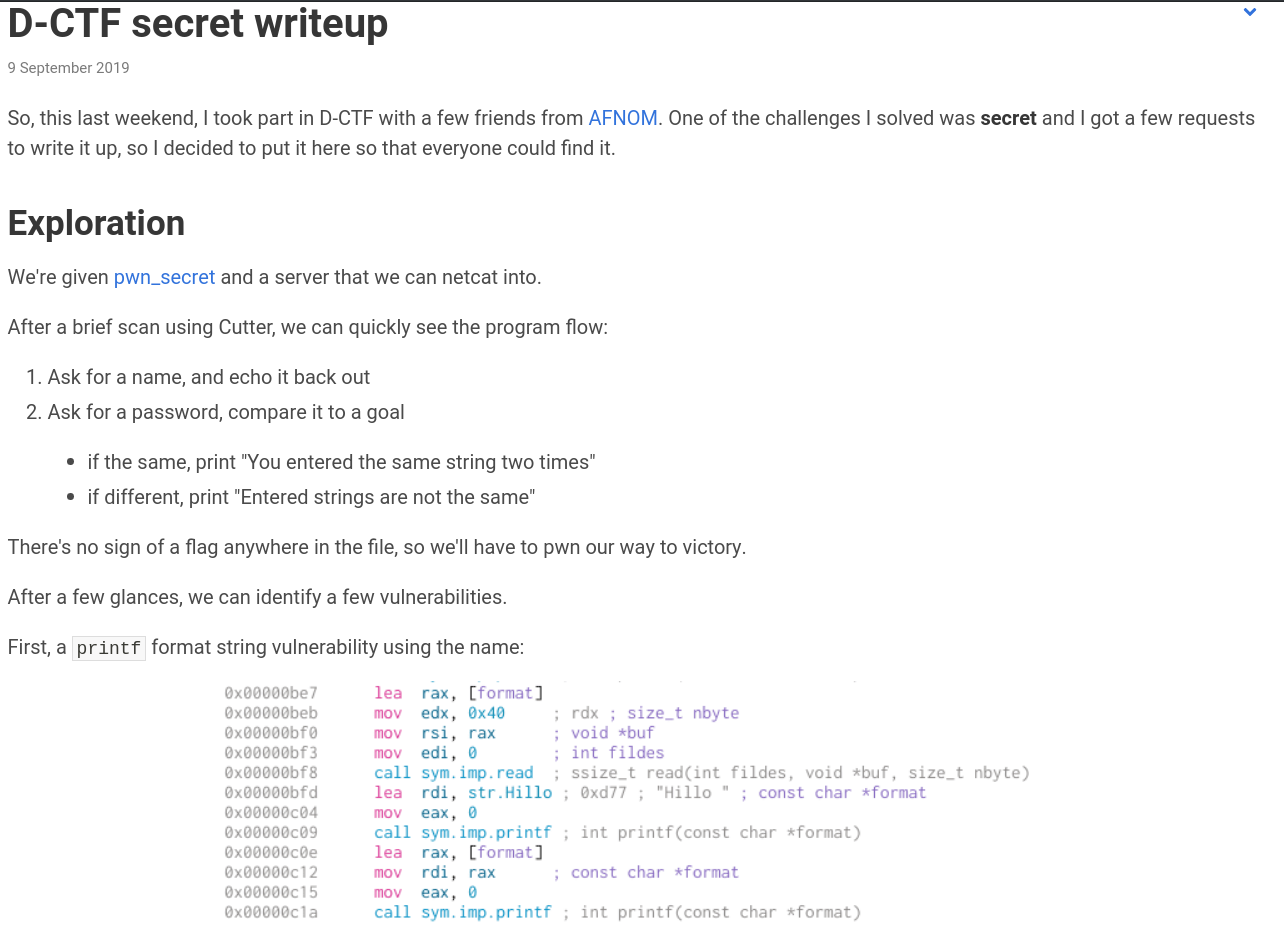}
    \end{minipage}

  \caption{\textbf{Left}: Learn \& Solve - Attendees work together to solve a challenge in an attempt to elicit the flag; \textbf{Right}: Share - One of the many online write-ups  (\url{https://jedevc.com/blog/dctf-secret-writeup/}).}
  \label{fig:learnsolveshare}
\end{figure}



\section{Comparison} \label{sec:comparison}
To aid with the generalisation of the two similar frameworks (hackathons and CTFs), and elicit important aspects of these learning environments it is important to consider insights from attendee feedback (Figures on pages~\pageref{fig:anal_1} and~\pageref{fig:anal_2}) 
- particularly the shared questions at \ac{HTM} 4.0 and \ac{WTCTF} 2020. Though this feedback comes from two specific implementations, or instances, of the frameworks, the lessons learned from each are still invaluable, and help with the abstraction of the underlying concepts. To this end, the responses to each of the seven statements (where X corresponds to either `\ac{HTM}' or `WhatTheCTF!?'),

\begin{enumerate}[label={Q}\arabic*.]
    \item I met new people at X;
    \item I felt welcome at X;
    \item I learned something new at X;
    \item I enjoyed X;
    \item X has inspired me to attend similar events;
    \item When I needed help, I was able to get it easily; and
    \item I got help from attendees as well as organisers; 
\end{enumerate}
on a 5-point Likert scale (Strongly Disagree; Disagree; Normal; Agree; Strongly
Agree) will be analysed, with (Q2) and (Q4), and (Q6) and (Q7) considered
jointly under broader headings.

\subsection{Q1: I met new people at X}

Although people are less likely to mingle outside of their team at \acp{CTF} than hackathons (likely due to their heightened competitive nature), this can both be alleviated (through enforced `no work' lunch breaks), and the general trend shows that both hackathons and \acp{CTF} offer effective environments for networking. Though in Higher Education this may serve a somewhat lesser purpose - as the cohort will likely already know one another quite well - this could help to form connections between groups of peers, and certainly encourages peer-mentoring at the events themselves. This is in large part down to the strong sense of community that is often fostered in these environments, and it is precisely this community feel and welcoming nature that lies core to the effective learning that they offer. This is corroborated in Q2 and Q7, which both highlight the incredibly supportive and inclusive spaces that both \ac{HTM} 4.0 and \ac{WTCTF} created.

\subsection{Q2/Q4: Engagement/Learning Environment}

As mentioned in Sections~\ref{sec:htm} and~\ref{sec:wtctf}, the organisers of these events dedicate significant amounts of time during the event to making the participants feel included, and making sure that no one leaves without learning new skills. As explored in the analysis of Q1 above, both hackathons and CTFs aim to foster welcoming environments for their participants. Whether through the normalisation of asking for help (both from `teachers' and peers), peer-mentoring (Q7), or a variety of breakout spaces and side events, they should work to encourage participants to actively engage with one another, and instil in them a sense of belonging. This, along with nurturing the eponymous idea of `hacking together' a functional yet imperfect solution, helps to curtail the significant Impostor Syndrome \citep{impostor} that many newcomers and `veterans' experience alike, and encourages the sharing of failure as well as success. This is a particularly pertinent issue in Higher Education in the form of reporting bias \citep{reporting_bias}. 

Student engagement is paramount to effective learning \citep{martin2018engagement, student_engagement}, and the attendee responses to Q4 are testament to the immense engagement that hackathons and CTFs can offer. Despite being learning environments, their departure from traditional lecture-based settings completely radicalises student perception of them. Notably, from a Transactional Analysis (TA) \citep{transactional_analysis} perspective, this represents a shift from Parent-Child transactions in lectures (and other similar settings) to Adult-Adult transactions - treating the students as equals, and eliciting more natural, candid, and unbounded discussion. 

\subsection{Q3: I learned something new at X}
Perhaps key to all learning environments (by definition) is that the students in them learn something new, and the responses to Q3 affirm that this was the case both at \ac{HTM} and at \ac{WTCTF}. The overwhelming majority of respondents not only agreed with - but strongly agreed with - the statement. This reinforces the perceived efficacy of these environments as pedagogic tools. 

\subsection{Q5: X has inspired me to attend similar events}
Learning is an ongoing process, and one that - especially as a student in Higher Education - lasts for a prolonged period of time. For a method of teaching to be effective, it must be able to consistently engage with students over this period. If students are deterred by a certain method, their engagement will be lessened, and thus, teaching becomes less effective on the whole. Responses show that not only are the vast majority of students engaged by this type of teaching (Q4), but they are also encourages or inspired to actively seek out similar learning opportunities for themselves (Q5). In the present day, it is quite common for undergraduate computer science students to be willing to travel to a hackathon or \ac{CTF} most weekends (alongside their studies) - both for the social aspects of the events, but also to learn new skills, build projects, and share what they eventually make (or don't make, as the case may be). Of course, this isn't to suggest that all students are, or possibly could be, this engaged, but by removing the barrier of distance - i.e. by running these sessions locally, similar to lectures, it seems likely that this engagement would be much higher.

\subsection{Q6/Q7: Help/Support}
A robust system of support for attendees is critical for both hackathons and \ac{CTF}s. Especially in the case that material or technologies are being largely self-taught during the event, attendees often find it hard to move past what may otherwise seem to be trivial obstacles - contributing significantly to feelings of Impostor Syndrome. It is therefore the role of the teacher(s) (organisers/volunteers) to mentor, and students (attendees) themselves to support one another. Without this support net of like-minded individuals in a similar situation, it is unlikely that as much material could be properly learned or applied in the intensely-short time frame over which such events often run. The responses to Q6 and Q7 show that most attendees at both \ac{HTM} and \ac{WTCTF} were not only able to access help when it was required, but for the most part, they were also able to access that support from their peers. This not only provides a more supporting and engaging environment for the students, but also equips them with important interpersonal skills, and helps to reinforce their own learning \citep{Snowden2019, reeves2019working}.

\begin{landscape}
\begin{figure}
\vspace{-2em}
  \centering

  \includegraphics[width=0.40\linewidth]{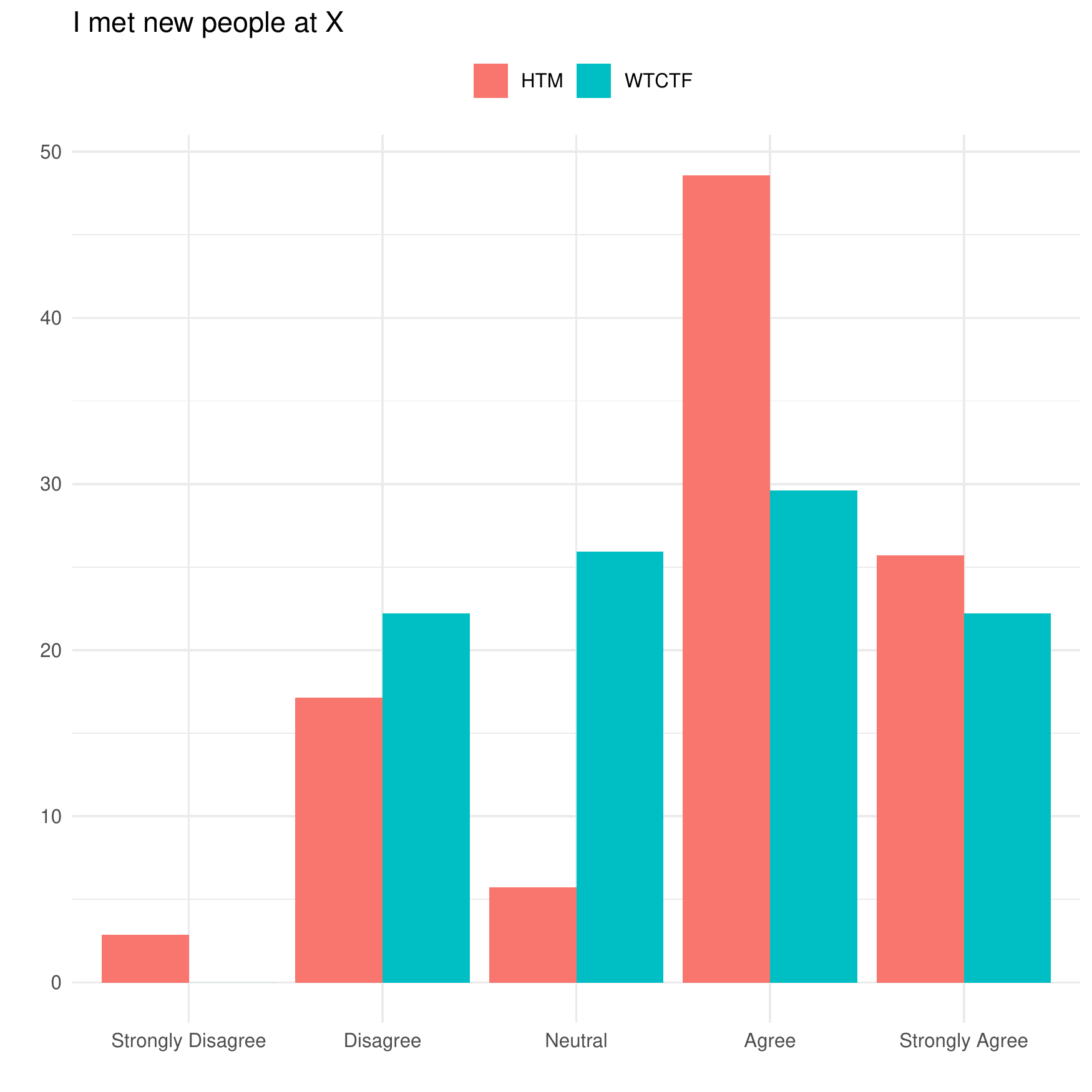}
  \includegraphics[width=0.40\linewidth]{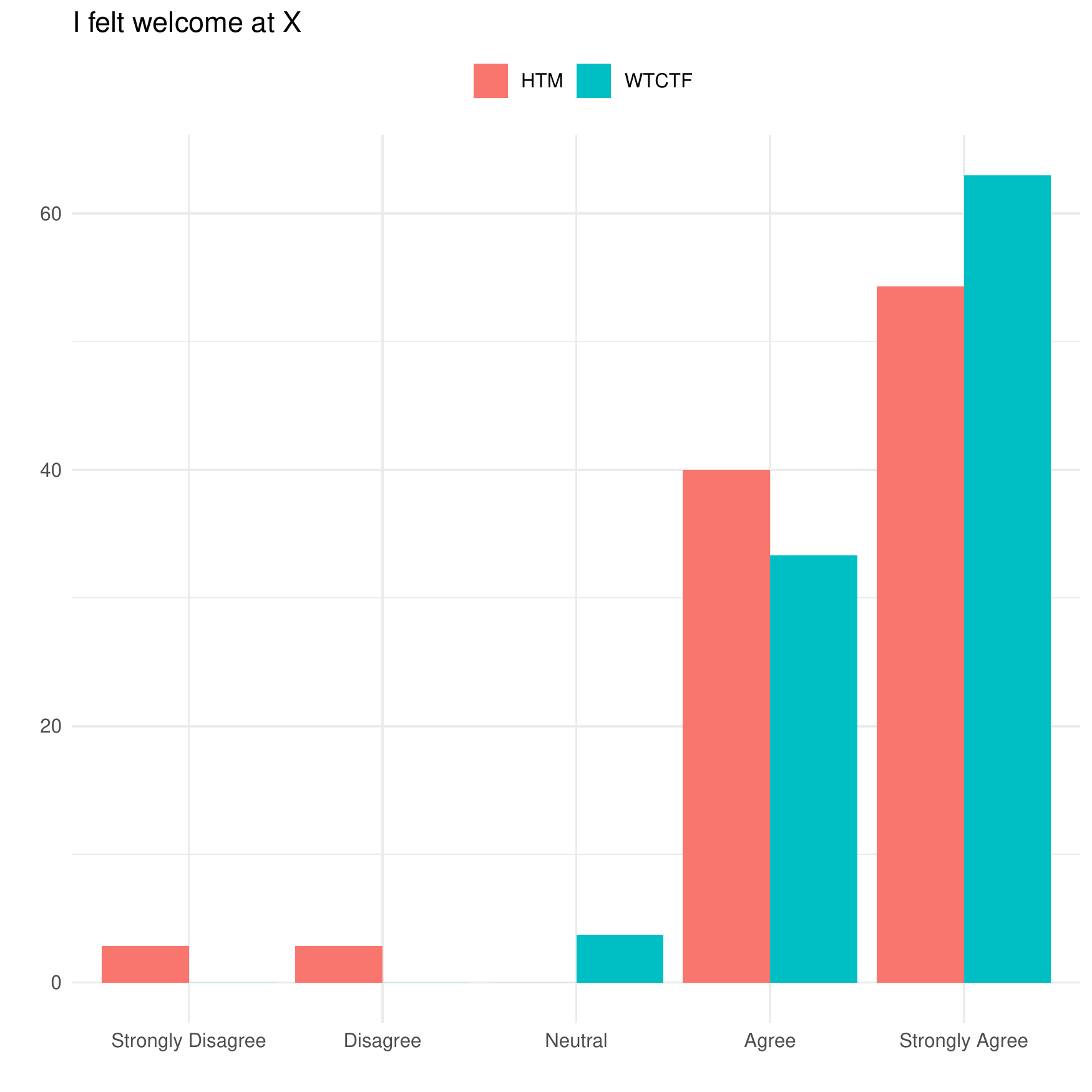}
 
  \includegraphics[width=0.40\linewidth]{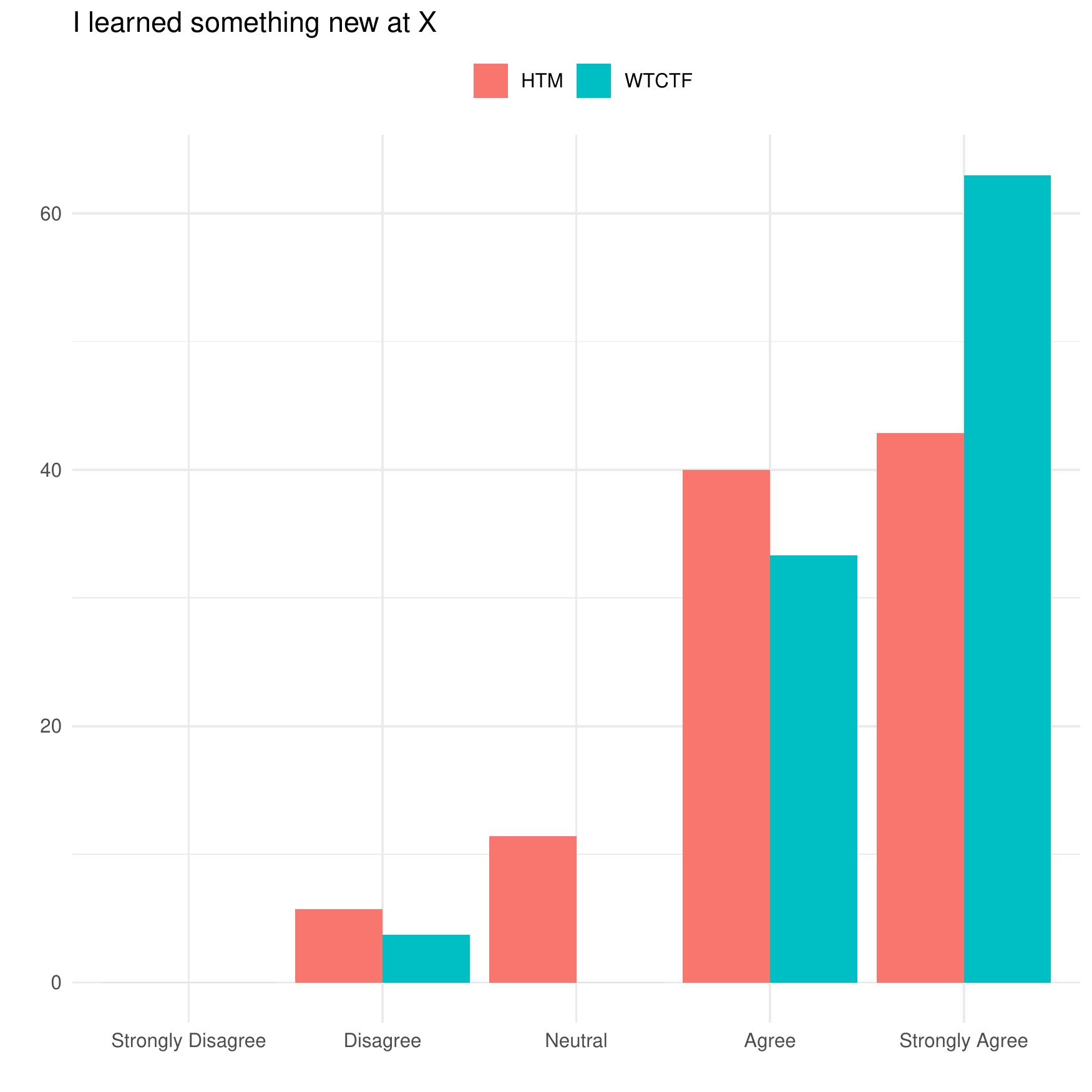}
  \includegraphics[width=0.40\linewidth]{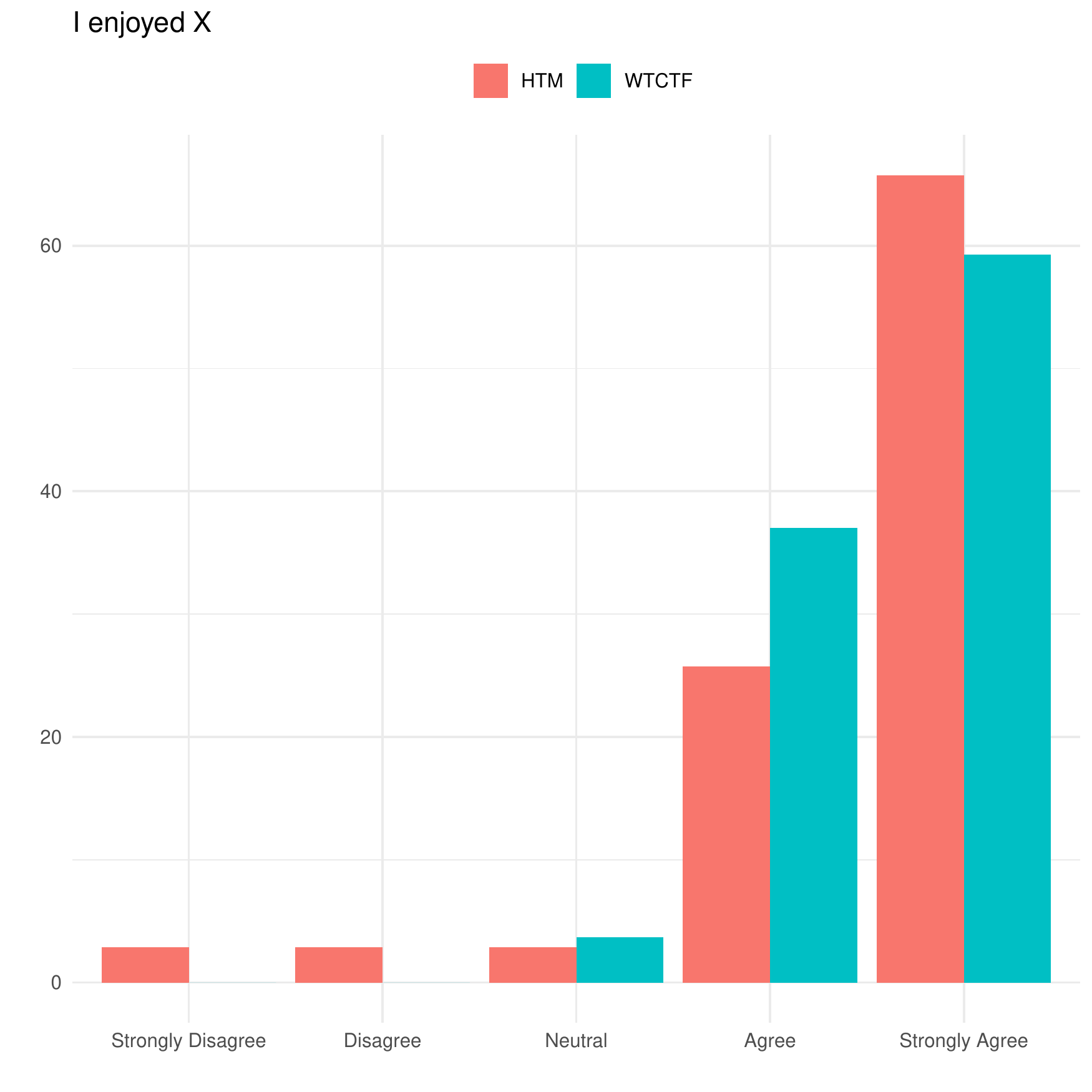}

  \caption{The first four (of seven) questions posed to participants.} 
 \label{fig:anal_1}
\end{figure}

\begin{figure}
  \centering
  \vspace{-2em}

  \includegraphics[width=0.40\linewidth]{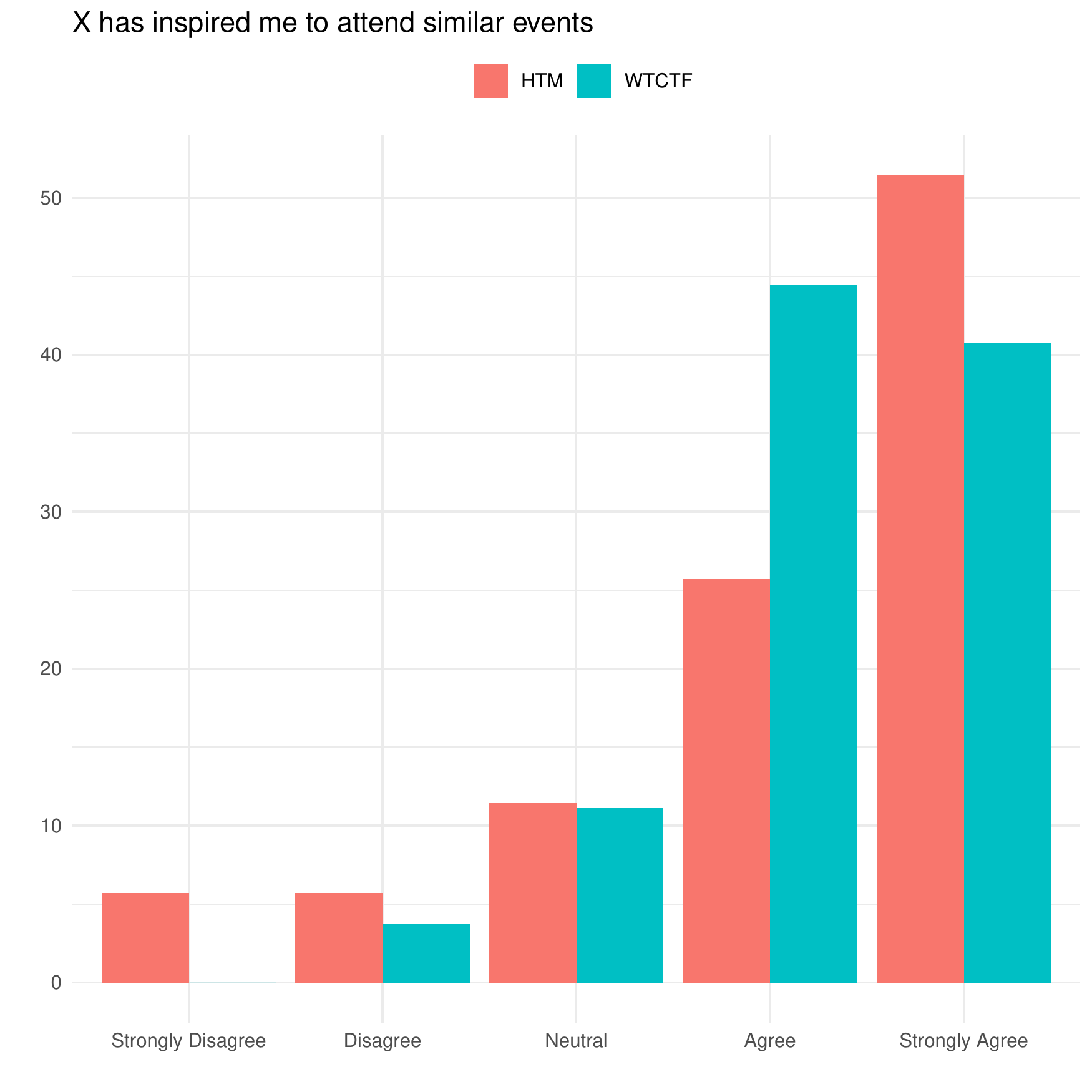}
  
  \includegraphics[width=0.40\linewidth]{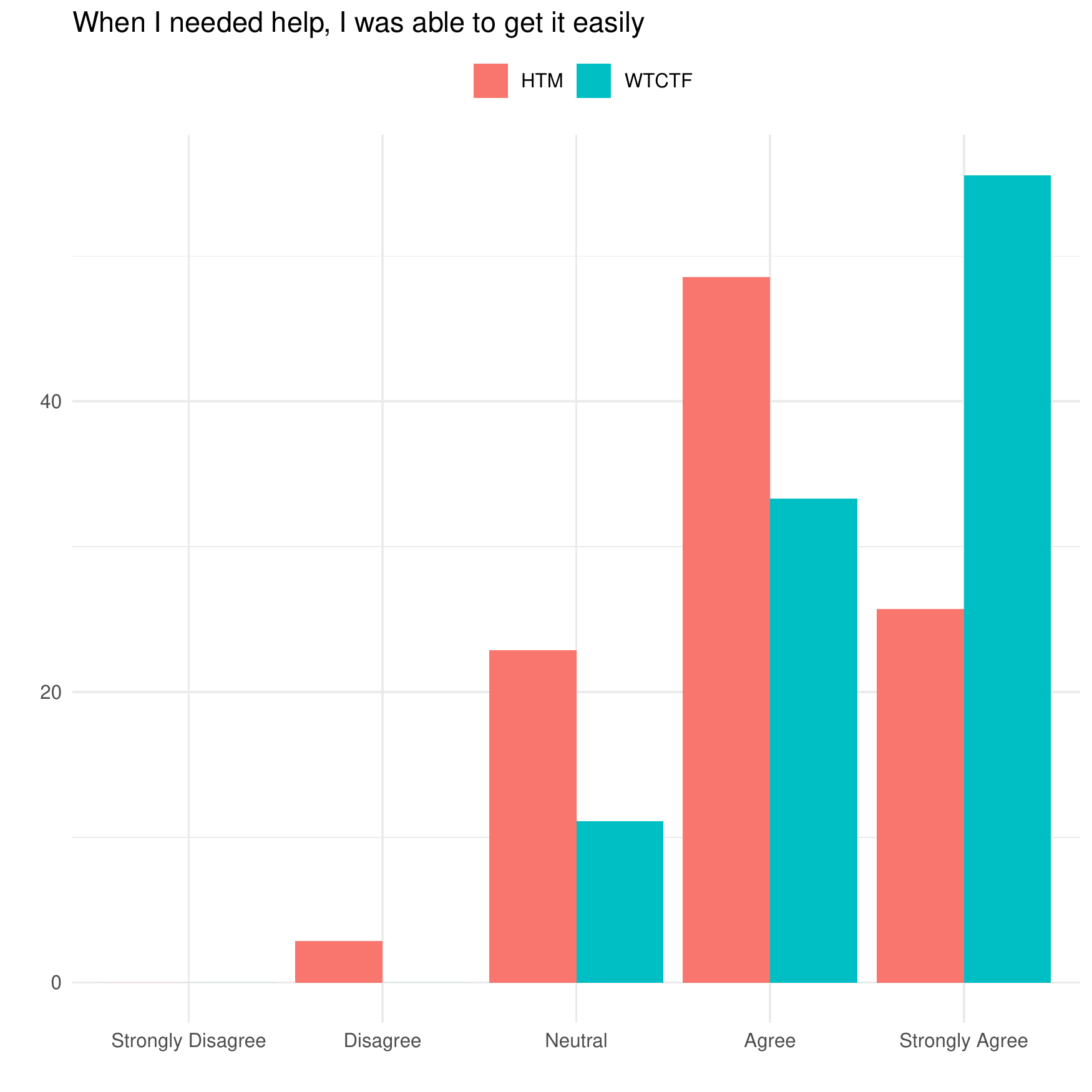}
  \includegraphics[width=0.40\linewidth]{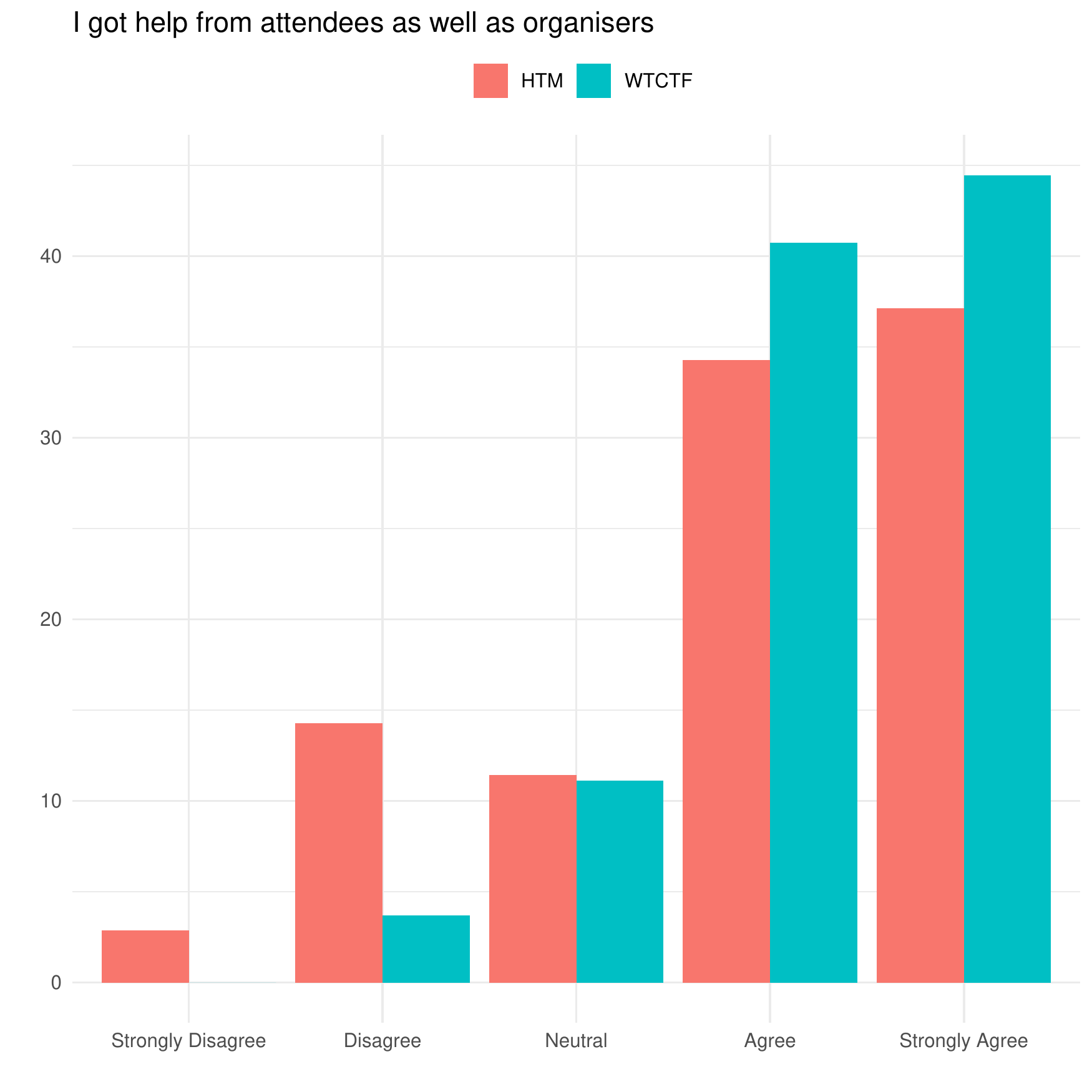}

  \caption{The final three questions posed to participants.} 
  \label{fig:anal_2}
\end{figure}
\end{landscape}

\section{Generalisation} \label{sec:generalisation}
\begin{figure}
    \centering
    \includegraphics[scale=0.45]{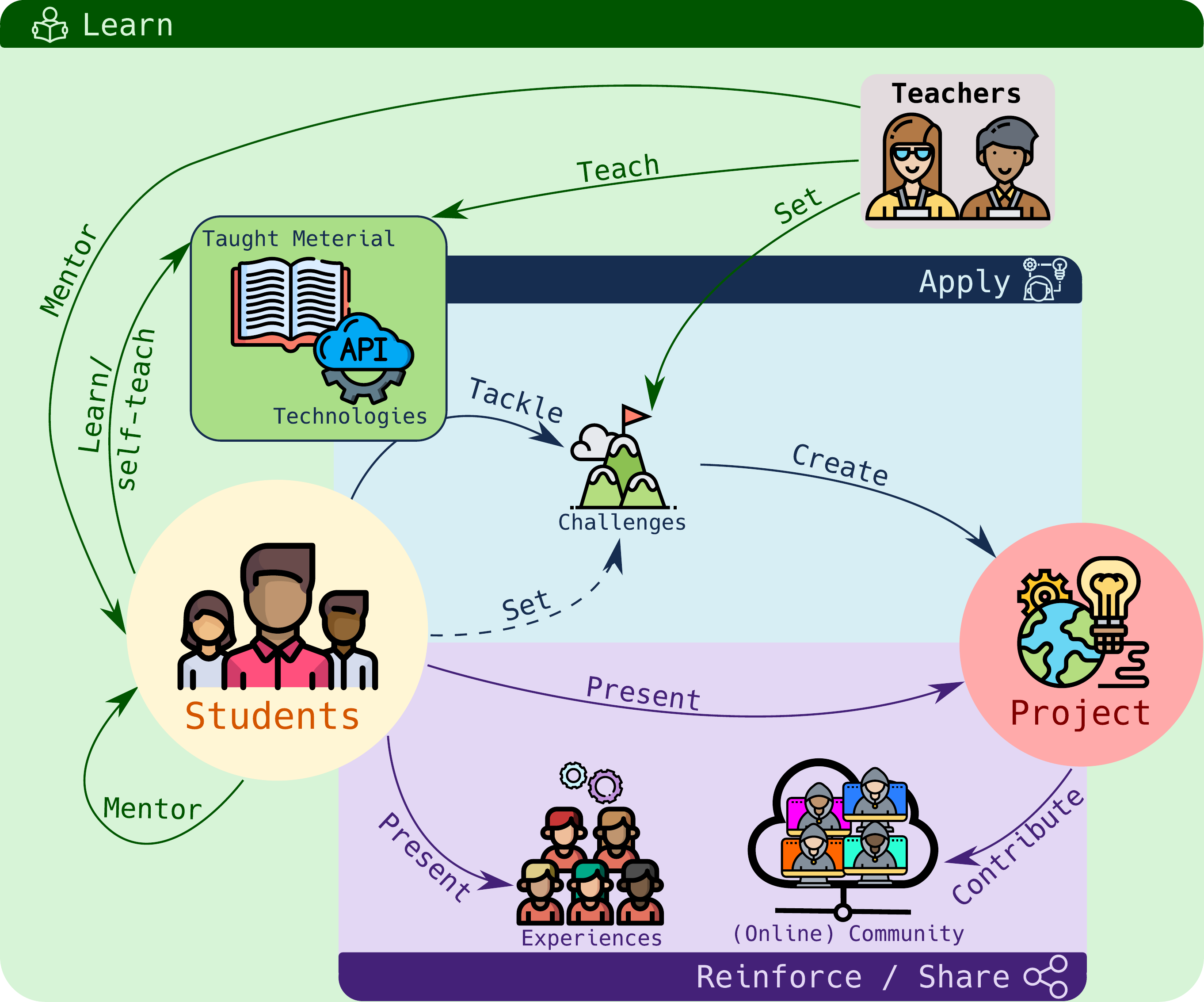}
    \caption{The generalised framework, centered around Learn-Apply-Reinforce/Share.
    Interactions between entities are depicted as arrows.}
    \label{fig:generalised}
\end{figure}

Though traditionally applied in a computer science (or STEMM) context, it is unequivocally the case that similar approaches can be utilised throughout Higher Education (and potentially further afield). By generalising the two frameworks (set out in Figures~\ref{fig:hackathon} and~\ref{fig:ctf}) for hackathons and \ac{CTF}s respectively, it is possible to consider an approach focused on three main stages - Learn, Apply, and Reinforce/Share (Figure~\ref{fig:generalised}). Here, a clear parallel can be drawn to the ``Learn, Build, Share" mantra of hackathons, or that of ``Learn, Solve, Share" for CTFs. This framework can be considered both in terms of a short, high-intensity `one-off' event, or as a project over a longer period of time. The importance of this distinction becomes clearer when considering distance learning (Section~\ref{sec:dl}). 

It is also important to consider these `phases' as largely asynchronous (with the exception of sharing work to the entire cohort). Students will regularly move between learning new material - especially that which is self-taught - and using that material to develop their project. Further, through peer mentoring, the `Reinforce' aspect is also prevalent throughout. This fluidity between stages results in a creation/development process very similar to the agile notion of Rapid Application Development (RAD) \citep{mackay2000reconfiguring} (albeit applied more broadly, and without a focus on software development), with the students rapidly prototyping aspects of their project (be that pieces of code, paragraphs, sections, or whatever), and using their team as a support network to gather and implement feedback iteratively. 

In the `Learn' part of the framework, students learn new materials, skills or technologies, that will be applicable to the challenges or problems that they'll work towards solving. This could be taught by the teacher(s) in a more traditional way, through talks or workshops ran by the teacher(s) or students, self-taught by the students, or in any other effective way. Further, peer-mentoring plays a significant part in the `Learn' and `Apply' stages, and should be encouraged. 

Within `Apply', the students take the knowledge and skills that they've gained, and use them to tackle a specific problem. This could be guided or set by the teacher(s) (similar to organisers and sponsors setting challenges in hackathons or CTFs), or the students could be left to define their own. Regardless, it's generally important to have a broad theme to unobtrusively guide their work. This initial broadness is necessary, so as to not stifle the creativity or imagination of the students. By effectively engaging the students as co-creators \citep{brand2019student}, it is possible to engage them much more effectively, as they're likely to be significantly more invested in creating and developing their project. One of the additional benefits around this self-definition of a problem (and thus, engagement with it), is a reduction in the urge to rationalise their creation post-hoc (i.e. fitting a solution to a problem), which in turn encourages students to focus on a much more goal-oriented approach. 

The `Reinforce/Share' stage encapsulates the presentation of students work to one another on an individual basis, but also to the whole cohort at the end of the learning. To many students, this may seem daunting, and in order to combat this, it is crucial to reassure them that what they have created need not be a full or perfect solution, and further, that they should talk about both their successes, but also their failures, and obstacles that they faced in their teams. By normalising this honesty and openness, the students are encouraged to learn from these failures, and often gain confidence by seeing that their peers all encountered challenges too. This can prove invaluable when tackling Impostor Syndrome within cohorts. Drawing on inspiration from CTFs, it is also positive to encourage students to create a write-up, not only of their project, but also of their experience and learnings. This can then be shared either online or offline, and in some cases, could contribute more widely to the community. 

Though not depicted in the diagram of the framework (for simplicity), feedback is universally important within it - and in all directions - that is,
\begin{itemize}
    \item Student $\to$ Student - through peer-mentoring and collaboration within teams;
    \item Teacher $\to$ Student - through guidance and mentoring;
    \item Student $\to$ Teacher - in the form of feedback about the learning environment;
    \item Teacher $\to$ Teacher - with the teacher acting as a reflective practitioner. 
\end{itemize}

Through these various feedback vectors, the learning environment (and learning experience of the students) is enhanced - especially when the Student $\to$ Teacher and Teacher $\to$ Teacher feedback is internalised, and implemented in future practice. 

\section{Using the Framework in a DL Context} \label{sec:dl}
Distance Learning (DL) is increasingly becoming a priority in the Higher Education sector, with increasing demand, and a variety of research being conducted into it \citep{shifting_dl, clark2020distance}. Particularly during the ongoing COVID-19 crisis, it has been brought to the forefront of university education, with institutions having to rapidly develop and deliver online teaching to millions of students worldwide. Within DL, as within regular learning, it is imperative to maintain student engagement, but the state of such engagement is significantly different within this context. Whereas it may be possible to engage students in high-intensity short events like hackathons when held in person, it is much harder to sustain that level of intense focus and engagement at distance. Thus, it is important to consider how best to alter the framework (or at least specific implementation of it) to better-suit these evolving needs.

The vast majority of CTFs are hosted online. In these cases, the onus of collaboration falls on the individual participants, who must form teams, and establish desired channels of communication with one another prior to the event. For the large part, the community aspect of these online CTFs comes both from the camaraderie within teams, and post-hoc from contributing to the broad online community through write-ups. This makes CTFs relatively well-adapted to distance learning, but does place a significantly more active role on participants - who must more proactively engage with the CTF in advance - to sign-up, form teams, decide on communication, and other tasks. When looking to transition the minority of CTFs that are held in-person to an online delivery, however, it may not be as simple as to follow the example set by their online counterparts - the lack of persistent community throughout the event would pose a fundamental change. To this end, additional provision should be made to help participants engage with one another, and to encourage and support typically underrepresented or disadvantaged participants to take part.

In contrast, the vast majority of hackathons are held in person. This is largely because of the aforementioned relative ease of engagement in that format, but also due to the underlying necessity for a community in which it is possible to easily seek help, share your ideas and work, and generally feel welcome. It is clearly not always feasible to run this kind of event in person, however - as highlighted quite pertinently by the ongoing pandemic. Thus, it is essential to adapt the approach such that it may effectively function online. When considering this kind of transition to online delivery (though not necessarily just in the context of hackathons), it is imperative to maintain the community. This requires a firm grasp of the appropriate technology for the job, as it is a relatively delicate art, and can prove easy to get wrong. Compared to community interactions in `real life', there are many more hidden obstacles that students may face when trying to engage with this online community, and if the resulting barrier to entry is too high, it is much more common for students to simply not engage.

One such tool is Discord\footnote{\url{https://discord.com/}} - a tool originally aimed towards gamers. Its intuitive design, extensibility (through bots), and flexibility make it a great candidate for cultivating the community required to underpin this kind of learning environment. On Discord, persistent (i.e. present 24/7) communities are called `Servers', which can be set up by anyone. Members can join these servers through an invite link (which makes them easy to disseminate, and reduces the barrier to entry), and engage with one another in both text channels, and voice/video channels. In addition, members can also be assigned `roles' such as `Teacher', `Student', or `Mentor', and these can be customised to grant various permissions. In addition, Discord have recently put together a magnificent step-by-step guide for teachers in all settings - helping them to set up virtual classrooms with ease \footnote{\url{https://blog.discord.com/how-to-use-discord-for-your-classroom-8587bf78e6c4}}.

\begin{figure}
    \centering
    \includegraphics[scale=0.35]{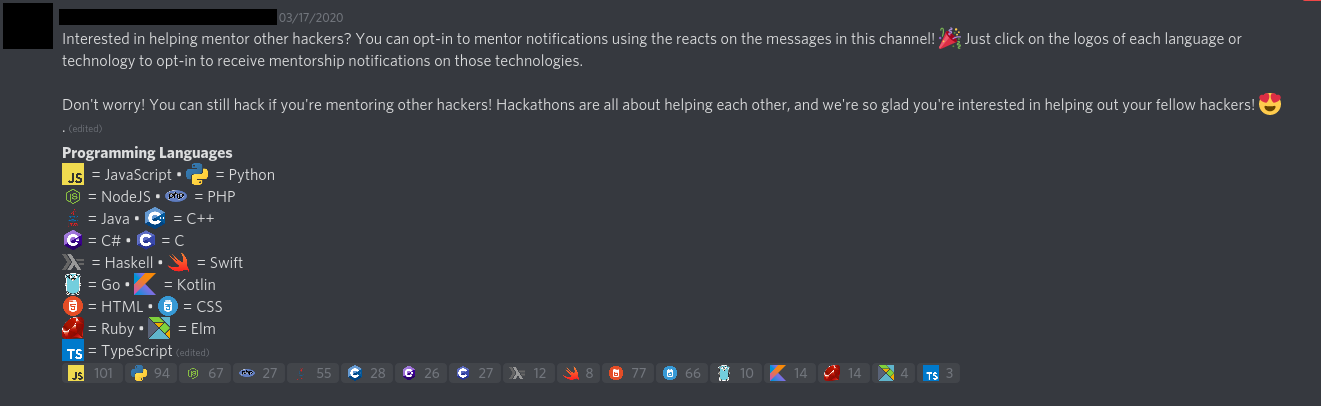}
    \caption{A snippet of HackQuarantine's mentoring system - allowing participants to self-define as mentors, and contact one another easily for help.}
    \label{fig:hq}
\end{figure}

Though Discord is sometimes used at hackathons such as HTM to supplement the in-person learning, they are much less-often used to facilitate one that is entirely online. In response to the COVID-19 pandemic, however, a hackathon called HackQuarantine \footnote{\url{https://hackquarantine.com/}} was set up and ran almost entirely on Discord. Key to much of this was their utilisation of an innovative mentoring system that not only allowed participants to quickly seek help, but also allowed them to self-define as a mentor for specific skills (in their case, programming languages). By using the Discord bot, `Zira' \footnote{\url{https://zira.gg/}}, they were able to automatically assign specific Discord roles to users that clicked on specific reactions (Figure~\ref{fig:hq}) under a pinned message. Each of these roles could then be specifically mentioned in the text chat when someone needs help (eg. @Python), which directly facilitated mentoring, and made it much easier for people to ask for help.

Thus, given the right technology, and a little momentum within a cohort, it is clear that the Learn-Apply-Reinforce/Share framework of learning is not only applicable to, but efficacious for, online delivery - with HackQuarantine engaging over 3500 participants, and `outputting' just shy of 250 projects \footnote{\url{https://hackquarantine.devpost.com/submissions/}}. One important factor to consider, however, is the time frame over which to run these `sessions'. Though in-person hackathons and CTFs tend to last between 8 and 48 hours, it is worth noting that this works largely due to the potential to create a much more focused environment. When students are working from home, for example, it is likely that there will be far more distractions (and thus a lack of focus).

This, coupled with the dispersion of students over a diverse range of time zones, means that the sessions must be run over a longer period of time. HackQuarantine, for example, ran over three weeks, which enabled students to better-engage with the learning. In order to maintain good levels of engagement, it is worth running `checkpoints' throughout this longer period, where students can informally share their current progress and obstacles or challenges with one another - interweaving the `Reinforce/Share' aspect of the framework, and providing a more active feedback loop for the students. Further, a range of `side events' will prove invaluable in maintaining this engagement, and can be used to mimic the breakout spaces provided at in-person events.







\newpage

\section{Conclusion \& Future Work} \label{sec:future}
This paper has presented a general pedagogic framework inspired by both hackathons and CTFs. Though these events are typically aimed at computer science or STEMM students, the framework is universally applicable. Split into three distinct sections, `Learn', `Apply', and `Reinforce/Share', it builds on a social constructivistic philosophy and a number of agile principles to create engaging and fast-paced learning environments with a  goal-oriented and project-based focus. The applicability of this framework in a distance learning capacity was also explored, with some best-practice technology suggested. 

Moving forwards, this framework shows promise, not just from a pedagogic standpoint, but also for research - with bodies such as the Alan Turing Institute hosting `Data Study Groups' \footnote{\url{https://www.turing.ac.uk/collaborate-turing/data-study-groups/}} - week-long `collaborative hackathons' for researchers. It could be possible to organise such events within individual departments or institutions to conduct and output small research projects or studies. Further, there is the potential for them to improve knowledge transfer between experienced and early-career researchers if the groups were formed of researchers at a variety of points in their career. 

To conclude, this broadly-applicable framework is currently of particular pertinence - both with growing demand for distance learning or bimodal delivery of material in the Higher Education sector, and in the midst of the COVID-19 pandemic. It is hoped that it will prove a useful resource for educators in Higher Education, and offer a wider outlook on potential tools and technologies to use in the current climate.

\section*{Disclosure}
Tom is one of two founders of HackTheMidlands, and was the `Science Lead' for HackQuarantine. Andreea runs WhatTheCTF!? as part of the AFNOM Ethical Hacking Society. 


\bibliographystyle{ieeetr}
\bibliography{lar_learning}

\end{document}